\documentclass[pra,twocolumn,tightenlines,showpacs,nofootinbib]{revtex4-1}
\usepackage{bm,dcolumn,amsmath,graphicx,amsfonts,amssymb}

\newcommand{\bra}[1]{\langle #1|}	
\newcommand{\ket}[1]{|#1\rangle}	
\newcommand{\threej}[6]{\begin{pmatrix}#1&#2&#3\\#4&#5&#6\end{pmatrix}}	
\newcommand{\sixj}[6]{\begin{Bmatrix}#1&#2&#3\\#4&#5&#6\end{Bmatrix}}	
\renewcommand{\v}[1]{\boldsymbol{#1}}		

\newcommand{\smallspace}{\rule{0pt}{3ex}}

\begin{document} 
\title{Strongly enhanced atomic parity violation due to close levels of opposite parity}
\author{B. M. Roberts}
\author{V. A. Dzuba}
\author{V. V. Flambaum}
\affiliation{School of Physics, University of New South Wales, Sydney, New South Wales 2052, Australia}
\date{ \today }

\begin{abstract}

We present calculations of nuclear-spin-dependent and nuclear-spin-independent parity violating amplitudes in Ba, Ra, Ac$^+$, Th and Pa.
Parity nonconservation in these systems is greatly enhanced due to the presence of very close electronic energy levels of opposite parity, large nuclear charge, and strong nuclear enhancement of parity-violating effects.
The presented amplitudes constitute several of the largest atomic parity-violating signals predicted so far.
Experiments using these systems may be performed to determine values for the nuclear anapole moment, a $P$-odd $T$-even nuclear moment given rise to by parity-violating nuclear forces.
Such measurements may prove to be valuable tools in the study of parity violation in the hadron sector.
The considered spin-independent transitions could also be used to measure the ratio of weak charges for different isotopes of the same atom, the results of which would serve as a 
test of the standard model and also of neutron distributions. 
Barium, with seven stable isotopes, is particularly promising in this regard.

\end{abstract}
\maketitle



\section{Introduction}

It is widely believed that the standard model is a low-energy manifestation of a more complete unified theory. 
Measurements of violations of fundamental symmetries in atoms, such as atomic parity nonconservation (PNC), provide a very effective channel for testing the standard model of elementary particles and for searching for new physics beyond it (see, e.g.~\cite{GFrev04,DzubaReview2012}).
Such studies complement measurements performed at high energy (e.g.~at CERN) for just a fraction of the cost.

Highly precise measurements~\cite{meas1,meas2} and calculations~
\cite{CsPNC89,CsPNC90,CsPNC92,CsPNC01,CsPNC02,CsPorsev,CsOur}
of the nuclear-spin-independent parity nonconservation amplitude in cesium have led to a determination of the nuclear weak charge that serves as the most precise atomic test of the electroweak theory to date. 
The result of this analysis 
is in reasonable agreement with the standard model~\cite{CsOur,CsQw}.
 However, it does indicate that further investigations in this field may lead to important new results. 
Furthermore, recent measurements made by the $Q_{\rm weak}$ Collaboration at the Jefferson Lab have led to the first determination of the weak charge of the proton~\cite{ProtonQw}. Combining this with the weak charge obtained via atomic parity violation in cesium leads to a value for the weak charge of the neutron.

It is unlikely that significant improvement in the measurements or calculations for cesium would be possible in the near future.
The trend in using atomic physics as a  probe for the low energy sector of the standard model is moving towards other possibilities~\cite{DzubaReview2012}.
For example, several proposals have been put forward to search for PNC in different atoms where either the theoretical or experimental accuracy may reasonably be expected to be better (e.g.~\cite{Ba+,Fortson,Yb+,Rb,FrLike,KVI,FrPNC}).
A promising alternative is to perform measurements of PNC in a number of different isotopes of the same atom~\cite{IsoChain}. 
This so-called `chain of isotopes' method requires no accurate atomic calculations. 
The ratio of the PNC signals for two isotopes does not depend on the electron structure. 
Here the accuracy is limited only by the knowledge of the (poorly understood) neutron distribution, see, e.g.~\cite{NeutronSkin,Brown2009}.

In this work we turn to another interesting area.
As well as the nuclear-spin-independent parity violating effect, the dominating effect in atoms that is caused by the interaction of the electrons with the weak charge of the nucleus, there are also spin-dependent effects, which arise from the interaction of the electrons with the so-called anapole moment of the nucleus~\cite{AnM,Flambaum1985}. 
A measurement of the anapole moment, a $P$-odd $T$-even nuclear moment that arises due to parity violating nuclear forces, would provide valuable information for the study of parity violation in the hadron sector~\cite{Khr91}.
As in the spin-independent case, the experiments require theoretical calculations for their interpretation. However, in the search for anapole moments the very high accuracy that is needed for the extraction of the weak charge is not required.
This frees up the possibility of exploiting favorable conditions found in more complicated atoms.

The ${}^{133}$Cs measurement by the Weiman group~\cite{meas1}  provides the only observation of a nuclear anapole moment. 
The quest for new measurements is also partly motivated by the requirement to perform an independent test of the existing cesium result in other systems. This is a  very important result and must be checked even if the accuracy is not improved. 
Moreover, the systems studied here have a very large enhancement in the PNC signal, which could make these systems even more favorable for observing the anapole moment.

Additionally, some isotopes of each of these atoms are believed to exhibit a very large nuclear enhancement of parity and time-invariance violating effects~\cite{ThPa1} (see also~\cite{ThPa2,ThPa3,CloseNuc84,CloseNuc83}).
Protactinium is a particularly interesting case in this regard, with a possibility of very close nuclear levels of opposite parity, which may lead to a huge enhancement in the PNC effects.
This is discussed in more detail in Section~\ref{sec:pa}.
Some isotopes of francium, radium and actinium also have close nuclear levels of opposite parity~\cite{ThPa2}.

In this work we provide calculations of both nuclear-spin-dependent  and nuclear-spin-independent  parity nonconserving effects that are enhanced by the presence of very close electronic levels of opposite parity.
The anapole moment induced PNC transition in neutral radium has been considered previously~\cite{FlambaumRa1999,DzubaRa2000}.
The PNC amplitude between the even ground and $^3D_2$  states was found to be more than $10^3$ times larger than the corresponding $6s-7s$ amplitude in cesium.
We revisit these calculations in Section~\ref{sec:ra}, improving the accuracy and verifying that the PNC signal is indeed greatly enhanced. 
In Sections~\ref{sec:ba} through \ref{sec:pa} we then proceed to calculate PNC, due both to the anapole moment and the nuclear weak charge, in neutral barium, singly-ionized actinium (an analogue of radium), neutral thorium, and protactinium. 

We believe the atoms and ions considered here are very promising candidates for experimental studies of parity violating nuclear effects. 
They may also be used to measure the ratio of weak charges in isotopic chain measurements.

\section{Theory}\label{sec:theory}

The Hamiltonian describing the parity violating electron-nucleus interaction can be expressed as the sum of the nuclear-spin-independent (SI) and nuclear-spin-dependent (SD) parts (using atomic units, $\hbar=|{e}|=m_e=1$, $c=1/\alpha\approx137$):
\begin{align}
	\hat h_{\rm PNC} &= \hat h_{\rm SI} + \hat h_{\rm SD} \notag\\
	&= \frac{G_F}{\sqrt{2}} \left(-\frac{Q_W}{2}\gamma_5 +   \frac{ \v{\alpha}\cdot \v{I}}{I} \varkappa \right)\rho(r),
\label{eq:hPNC}
\end{align}
where $G_F\approx 2.2225\times10^{-14}$ a.u. is the Fermi weak constant, $Q_W$ is the nuclear weak charge, $\v{\alpha}=\gamma_0\v{\gamma}$ and $\gamma_5= i \gamma_0\gamma_1\gamma_2\gamma_3$ are Dirac matrices,  $\v{I}$ is the nuclear spin and $\rho(r)$ is the normalized nuclear density, $\int \rho\, {\rm d}^3 r=1$. 

The strength of the SD-PNC interaction is proportional to $\varkappa$, a dimensionless coupling constant \cite{NoteDefinition}.
The SD-PNC interaction can be expressed as the sum of its three main contributions \cite{GFrev04}:
\begin{equation}
	\varkappa = \frac{K}{I+1}\kappa_{\rm a} - \frac{K-1/2}{I+1}\kappa_{Z} + \kappa_Q,
\end{equation}
where $K=(I+1/2)(-1)^{I+1/2-l}$ with $l$ the orbital angular momentum of the unpaired nucleon.
The dominating contribution in heavy atoms comes from $\kappa_{\rm a}$, the nuclear anapole moment~\cite{AnM}.
$\kappa_{Z}$ quantifies the contribution from the spin-dependent electron-nucleus weak interaction ($Z^0$ exchange)~\cite{Flambaum1977},
and $\kappa_Q$ is from the combination of the SI-PNC contribution ($Q_W$) with the magnetic hyperfine interaction~\cite{Flambaum1985} (see also \cite{Bouchiat1991,Johnson2003}).
For more information see, e.g., the review~\cite{GFrev04} and the book~\cite{Khr91}.

To lowest order, the nuclear weak charge is given in the standard model as 
\begin{equation}
	Q_W = -N + Z(1-4\sin^2\theta_{\rm W}). 
\end{equation}
Here $N$ and $Z$ are the number of neutrons and protons in the nucleus respectively, and $\sin^2\theta_{\rm W}\approx0.23$ is the Weinberg electroweak mixing angle \cite{PDG13}.

The interaction of the valence electrons with both the anapole moment and the weak charge of the nucleus leads to mixing between electronic states of opposite parity.  
This has the effect of allowing non-zero $E1$ transition amplitudes between states of equal parity.
Unlike the spin-independent PNC interaction, however, interaction with the anapole moment can mix electronic states with a change in total electron angular momentum $\Delta J=1$ (as well as $\Delta J=0$), and is dependent on nuclear spin, which means contributions from different hyperfine components are different.

The amplitude of a parity invariance violating $E1$ transition between two states of the same parity can be expressed via the sum over all opposite parity states $n$:
\begin{align}
	E_{\rm PNC}^{a\to b} &= \sum_n \Bigg[\frac{\bra{b}\hat d_{E1} \ket{n} \bra{n}\hat h_{\rm PNC}\ket{a}}{\epsilon_a - \epsilon_n} \notag\\
	&\phantom{= \sum_n \Bigg[}+
\frac{\bra{b}\hat h_{\rm PNC} \ket{n} \bra{n}\hat d_{E1}\ket{a}}{\epsilon_b - \epsilon_n} \Bigg],
\label{eq:fullsummation}
\end{align}
where $a$, $b$ and $n$ are the many-electron wavefunctions of the system in question, $\hat d_{E1}$ is the electric dipole transition operator,
$\hat h_{\rm PNC}$ is the operator of the parity violating interaction that gives rise to the transition, 
and the sum runs over all states of opposite parity. 
Here, $\ket{a}\equiv\ket{J_aF_aM_a}$ with $\v{F}=\v{I}+\v{J}$ the total atomic angular momentum.
Formulas linking equation~(\ref{eq:fullsummation}) to the reduced matrix elements of the relevant operators are given in the appendix.

There are several factors which contribute to the enhancement (or suppression) of the parity-violating signal in atomic transitions.
The first, pointed out by the Bouchiats~\cite{Bouchiat}, is that the PNC amplitude should scale a little faster than $Z^3$ ($Z$ the atomic number). 
For this reason it is natural to expect larger amplitudes in heavy systems.
Also, as is clear from Eq.~(\ref{eq:fullsummation}), the existence of close energy levels of opposite parity has the potential to produce a very large enhancement.
It is with these motivations in mind that we pursue large PNC signals in the heavy elements chosen for this work. 
The transitions studied here have opposite parity levels with energy intervals of $\sim10~{\rm cm}^{-1}$. 
For comparison, the energy gap for the largest contributing term to the $6s-7s$ PNC transition in Cs is $\sim10^4~{\rm cm}^{-1}$.

Perhaps the most important and hardest to predict, the final factor is the size of the weak-interaction matrix element between the opposite parity states. 
The most significant contribution to this comes from $s$-$p_{1/2}$ mixing~\cite{Khr91}.
Finding large atomic systems with close pairs of opposite parity levels is comparatively simple, however determining the extent of single-electron $s$-$p_{1/2}$ mixing generally requires complicated calculations. In the heavy atoms studied here this tends to suppress the final amplitude -- e.g.~there is a $10^3$ factor enhancement from the proximity of opposite parity states, but this does not necessarily transform directly to a $10^3$ enhancement in the amplitude.
As well as this, of course, is the fact that the SI interaction cannot mix states of different total angular momentum, and the SD interaction can only mix states with $\Delta J=0,1$ (and $J \neq 0 \to 0$).

Actually, the $s$-$p_{3/2}$ and $p_{3/2}$-$d_{3/2}$ weak mixing is not insignificant. 
This is mainly due to core polarization, without which these contributions would be practically zero.
This has the benefit of counteracting the suppression due to limited $s$-$p_{1/2}$ mixing, however it makes the calculations very sensitive to the usually smaller corrections such as correlations and core polarization. This makes determining the accuracy particularly difficult, especially in cases for which the amplitudes are small. 
If there is only a small amount of $s$-$p_{1/2}$ mixing, then the amplitude becomes very sensitive to core polarization, and is thus not particularly accurate, even if $E1$ amplitudes and energies are reproduced well. 
 In all cases, the PNC matrix elements are sensitive to configuration mixing.

\section{Calculations}

Starting with the relativistic Hartree-Fock (RHF) method with a  
$V^{N-M}$ potential \cite{Dzuba2005}, where $N$ is the total number of electrons and $M$ is the number of valence electrons,
 we make use of the combined Configuration Interaction (CI) and many-body perturbation theory (MBPT) method developed in Ref.~\cite{CI+MBPT1996}. 
For more detail on this method see also Refs.~\cite{Dzuba2011Yb,YbPolariz,CI+MBPT1998}.

Interactions with external fields and core polarization are taken into account using the time-dependent Hartree-Fock (TDHF) method, see e.g.~\cite{rpa1,rpa2}.
Note that we do not take into account the effect of core polarization due to simultaneous action of the weak and $E1$ fields. This is because we focus on the spin-dependent amplitudes for which the accuracy of the analysis is less important.
This `double-core-polarization' effect was the study of our recent work Ref.~\cite{DCP}.


The effective CI+MBPT Hamiltonian for the system of $M$ valence electrons has the form:
\begin{equation}
\hat H^{\rm eff} = \sum_i \hat h_1(r_i) + \sum_{i<j} \hat h_2(r_i,r_j),
\label{eq:hci}
\end{equation}
where $\hat h_1$ is the single-electron part of the RHF Hamiltonian,
\begin{equation}
 \hat h_1 =  c \v{\alpha}\cdot\v{\hat p} + c^2(\beta -1) - V^{\rm nuc.}+U^{\rm HF} + \hat\Sigma_1,
\label{eq:h1}
\end{equation}
and $\hat h_2$ is the two-electron part,
\begin{equation}
 \hat h_2(r_i,r_j) = \frac{1}{|\v{r_i}-\v{r_j}|} + \hat\Sigma_2(r_i,r_j),
\label{eq:h2}
\end{equation}
and the sum runs over the $M$ valence electrons.
In the above equations, $\v{\alpha}$ and $\beta$ are the Dirac matrices, $V^{\rm nuc.}$ is the nuclear potential (we use a Fermi-type nuclear charge distribution to calculate $V^{\rm nuc.}$), and $U^{\rm HF}$ is the RHF potential created by the $N-M$ electrons of the closed-shell core.
The additional term, $\hat\Sigma$, is  the correlation potential, without which these equations would correspond to the conventional CI method.
The correlation potential is used to take into account core-valence correlations (see Refs.~\cite{CI+MBPT1996,CI+MBPT1998} for details).
The single electron correlation potential $\hat\Sigma_1$ represents the interaction of a single valence electron with the atomic core, and $\hat\Sigma_2$, a two-electron operator, represents the screening of the valence-valence Coulomb interaction by the core electrons.

We calculate the correlation potential $\hat \Sigma_1$, which includes a summation of dominating diagrams including screening of the Coulomb interaction and the electron-hole interaction, 
to all orders of perturbation theory using relativistic Green's functions and the Feynman diagram technique~\cite{CsPNC89,CPM2}.  To construct the complete set of single-electron orbitals we use the B-spline technique~\cite{Bspline}.

As a test of the overall accuracy of the calculations, we also calculate the correlation potential  to only second-order in perturbation theory, $\hat \Sigma_1^{(2)}$.
The difference between calculations using $\hat \Sigma_1$ and $\hat \Sigma_1^{(2)}$ give a good indication of the uncertainty due to missed higher order correlation corrections.


In the evaluation of the matrix elements, the operators $\hat{d}_{ E1}$ and $\hat{h}_{\rm PNC}$
are modified to include the effect of the polarization of the core electrons due to interaction with the external $E1$ and weak fields: 
\begin{equation}
\begin{gathered}
\hat{d}_{ E1} \rightarrow \hat{d}_{ E1} + \delta V_{ E1},\\
\hat{h}_{\rm PNC} \rightarrow \hat{h}_{\rm PNC} + \delta V_{\rm PNC}. 
\end{gathered}
\end{equation}
Here $\delta V_{ E1}$ ($\delta V_{\rm PNC}$) is the modification to the RHF potential due to the effect of the external field $\hat{d}_{E1}$ ($\hat h_{\rm PNC}$). 
In the TDHF method, the single-electron orbitals are perturbed in the form $\psi=\psi_0+\delta\psi$ where $\psi_0$ is an eigenstate of the RHF Hamiltonian, and $\delta\psi$ is the correction due to the external field.
The corrections to the potential are then found by solving the set of self-consistent TDHF equations for the core states:
\begin{equation}
(\hat H_0 -\varepsilon_c)\delta \psi_{c} = -(\hat f +\delta V_f)\psi_{0c},
\label{eq:RPA}
\end{equation} 
where the index $c$ denotes core states and $\hat f$ is the operator of external field (be that $\hat{d}_{E1}$ or $\hat h_{\rm PNC}$).
We also use this method to compute the sum Eq.~(\ref{eq:fullsummation}).

Calculation of the PNC amplitude requires a summation of the complete set of states.
We use the Dalgarno-Lewis method~\cite{DalgarnoLewis} to perform the summation. 
In this method the amplitude is reduced to
\begin{equation}
E^{a\to b}_{\rm PNC} = \bra{\delta\psi_b}\hat d_{E1} +\delta V_{E1}  \ket{\psi_a} + \bra{\psi_b}\hat d_{E1} +\delta V_{E1}  \ket{\delta\psi_a},
\end{equation}
where $\delta\psi_n$ is the correction to the wavefunction found by solving Eq.~(\ref{eq:RPA}) with $\hat f=\hat h_{\rm PNC}$ for the relevant valence states.

\subsection*{Calculating PNC with the resonant term}


Radium, barium and singly-ionized actinium have two valence electrons above a 
closed-shell core. For these relatively simple systems the above method works quite well.
We generate the wavefunctions and energies using a full CI calculation allowing double valence excitations, with core-valence correlations taken into account as described above. 
For thorium and protactinium, with more than two valence electrons, we use slight variations of the above method (discussed in later sections) and do not try to compute the entire sum.
Due to the presence of the very close opposite parity levels, the transitions in question have a single dominating term, contributing upward of 95\% to the total amplitude. 
For this reason it is a good first approximation to calculate this term alone. To do this we calculate the relevant matrix elements of the $E1$ and weak interactions including core polarization, and use the experimental energy difference to compute the term.

For Ba, Ra and Ac$^+$ we perform the entire summation to determine the whole amplitude.
A problem that occurs though is that the existence of the close levels makes this method numerically unstable.
Even if the energy levels are computed to very high accuracy, the relevant energy interval may be very wrong.
For example, the experimental energy gap between the even $^3D_2$ and the odd $^3P_1$ levels in radium is 5.41 cm$^{-1}$. 
Our calculations for the energies of these states vary from experiment by just 5\% and 1\% respectively, however, we calculate this interval to be 828 cm$^{-1}$. 
This would lead to an error of several orders of magnitude.
There are two methods we can use to remedy this.

The first and simplest method is to rescale the single-electron correlation potential,  
 i.e.~$\hat\Sigma_1\to\lambda\hat\Sigma_1$ in Eq.~(\ref{eq:h1}).
Different parameters are used for each partial wave ($s$, $p$ etc.) and are chosen to reproduce the relevant experimental energy interval exactly.
 For radium, the ionization energy of the the ground state was also fitted to match exactly with experiment.
It is worth noting that these scaling parameters are close to unity, indicating the already reasonable accuracy. 
For Ra the parameters chosen were $\lambda_s=0.994$, $\lambda_p=1.046$, and $\lambda_d=0.893$.
For Ac$^+$ they were $\lambda_s=0.957$, $\lambda_p=1.016$ and $\lambda_d=0.917$, and for 
Ba  they were  $\lambda_s=1.010$, $\lambda_p=0.897$, and $\lambda_d=0.933$.

The other approach is not to perform any re-scaling of $\Sigma$, but to use orthogonality conditions in the summation to extract the dominating term, and rescale it by a factor $\Delta E^{\rm Calc.}/\Delta E^{\rm Exp.}$.
To do this, we force the intermediate states in Eq.~(\ref{eq:fullsummation}) to be orthogonal to the state causing the dominating term and perform the summation. By comparing the results from this with the results of the summation without the orthogonality enforced we can separate this `main' term from the sum and proceed to re-scale it.
If good agreement exists between these two methods as well as with calculating the matrix elements directly it is indicative of good numerical accuracy, and this is what we find.


  \begin{table}%
    \centering%
    \caption{Comparison of calculated energy levels with experiment (Ref.~\cite{NIST}) for Ba, Ra and Ac$^+$. 
Units are cm$^{-1}$.} 
\begin{ruledtabular}%
  \begin{tabular}{lllrrr}
 & \multicolumn{2}{c}{}  & \multicolumn{2}{c}{Calc.} & \multicolumn{1}{c}{} \\
\cline{4-5}\smallspace
Atom & \multicolumn{2}{c}{State}  & \multicolumn{1}{c}{$\Sigma$}& \multicolumn{1}{c}{$\lambda\Sigma$} & \multicolumn{1}{c}{Exp.} \\
\hline   
\smallspace
Ba&$6s^2$   &  ${}^1S_0$  &  0  &  0  &  0  \\
&$5d6s$   &  ${}^3D_1$  &  8180  &  8684  &  9034  \\
&   &  ${}^3D_2$  &  8368  &  8865  &  9216  \\
&   &  ${}^3D_3$  &  8765  &  9243  &  9597  \\
&$5d6s$   &  ${}^1D_2$  &  10772  &  11309  &  11395  \\
&$6s6p$  &  ${}^3P_0$  &  12387  &  12677  &  12266  \\
&   &  ${}^3P_1$  &  12748  &  13031  &  12637  \\
&   &  ${}^3P_2$  &  13617  &  13877  &  13515  \\
&$6s6p$  &  ${}^1P_1$  &  17737  &  18080  &  18060  \\
&$5d^2$  &  ${}^3F_2$  &  19669  &  20605  &  20934  \\
&   &  ${}^3F_3$  &  20007  &  20928  &  21250  \\
&   &  ${}^3F_4$  &  20409  &  21314  &  21624  \\
&$5d6p$   &  ${}^3F_2$  &  21242  &  22015  &  22065  \\
&   &  ${}^3F_3$  &  22121  &  22866  &  22947  \\
&   &  ${}^3F_4$  &  22955  &  23675  &  23757  \\
&$5d2$  &  ${}^1D_2$  &  22216  &  23062  &  23062  \\
&$5d6p$   &  ${}^1D_2$  &  22320  &  23074  &  23074  \\
&$5d^2$  &  ${}^3P_0$  &  22086  &  22949  &  23209  \\
&   &  ${}^3P_1$  &  22340  &  23199  &  23480  \\
&   &  ${}^3P_2$  &  22895  &  23727  &  23919  \\
\smallspace
 Ra        & $7s^2$ & $^1S_0$ & 0 & 0    & 0 \\
  &  $7s7p$  &  $^3P_0$  &  13285  &  13102  &  13078  \\
  &    &  $^3P_1$  &  14170  &  13999  &  13999  \\
  &    &  $^3P_2$  &  16835  &  16694  &  16689  \\
  &  $6d7s$  &  $^3D_1$  &  13079  &  13756  &  13716  \\
  &    &  $^3D_2$  &  13342  &  13994  &  13994  \\
  &    &  $^3D_3$  &  14067  &  14642  &  14707  \\
  &  $6d7s$  &  $^1D_2$  &  16742  &  17318  &  17081  \\
  &  $7s7p$  &  $^1P_1$  &  20487  &  20432  &  20716  \\
  &  $7s8s$  &  $^3S_1$  &  26691  &  26658  &  26754  \\
\smallspace
Ac$^+$        & $7s^2$ & $^1S_0$ & 0& 0     & 0 \\
  &  $6d7s$  &  $^3D_1$  &  3917  &  4355  &  4740  \\
  &     &  $^3D_2$  &  4406  &  4836  &  5267  \\
  &     &  $^3D_3$  &  6579  &  6911  &  7427  \\
  &  $6d7s$  &  $^1D_2$  &  8403  &  8886  &  9088  \\
  &  $6d^2$  &  $^3F_2$  &  12023  &  12849  &  13236  \\
  &     &  $^3F_3$  &  13762  &  14557  &  14949  \\
  &     &  $^3F_4$  &  15644  &  16281  &  16757  \\
  &  $6d^2$  &  $^3P_0$  &  16250  &  17039  &  17737  \\
  &     &  $^3P_1$  &  17530  &  18290  &  19015  \\
  &     &  $^3P_2$  &  21615  &  22199  &  22199  \\
  &  $6d^2$  &  $^1D_2$  &  18053  &  18773  &  19203  \\
  &  $6d^2$  &  $^1G_4$  &  20692  &  20804  &  20848  \\
  &  $7s7p$  &  $^3P_0$  &  21453  &  21048  &  20956  \\
  &     &  $^3P_1$  &  22550  &  22181  &  22181  \\
  &     &  $^3P_2$  &  28612  &  28328  &  26447  \\
  \end{tabular}%
\end{ruledtabular}%
    \label{tab:energy}%
  \end{table}%

In Table~\ref{tab:energy} we present our calculated energy levels for Ba, Ra and Ac$^+$ along with experimental values for comparison.
We present calculations using both the unscaled correlation potential (column $\Sigma$) as well as the calculations including the rescaled potential (column $\lambda\Sigma$). 
Here we have chosen the scaling parameters to reproduce the energies of the relevant close opposite parity levels, as opposed to with achieving good overall accuracy in mind.
The unscaled energies for barium and radium are already very good, and in most cases this scaling improves the accuracy.

For Ac$^+$ the agreement with unscaled energies is not as good, however it should be noted that the intervals between levels are reproduced to a much better accuracy than the levels themselves, indicating most of the error is probably associated with determining the ground state energy. In this case the rescaling improves the accuracy for all levels.

\subsection*{Testing the method and accuracy}\label{subsec:yb} 

Ytterbium, like Ba, Ra and Ac$^+$ has two valence electrons above a closed shell core.
The parity violating $6s^2\,{}^1S_0\to6s5d\,{}^3D_1$ transition has contributions from both the spin-independent and spin-dependent parts of the PNC Hamiltonian, and is enhanced by the proximity of  the odd $6s6p\,{}^1P^o_1$ level to the upper ${}^3D_1$ level in the transition.
The energy interval between these levels is just 579 cm$^{-1}$. Though not as small as the other intervals studied in this work, it still means this transition in Yb has a large dominating term, contributing more than 80\% to the total amplitude.
Several calculations exist in the literature for parity nonconservation in neutral ytterbium~\cite{Dzuba2011Yb,demille,porsev-yb-si,das-yb-si,porsev-yb-sd,das-yb-sd}.
Therefore studying PNC in this transition for Yb will serve as a useful test for the method.

In Table~\ref{tab:compare-yb} we present calculations of both the spin-independent and spin-dependent parts of the PNC amplitude for this transition in ytterbium, and compare these values with those calculated in several other works.
We present calculations using both methods described above, that is using a scaled correlation potential ($\lambda\Sigma$) that is chosen to reproduce exactly the energy interval of the dominating term, and also using the orthogonality conditions to extract the dominating contribution and then re-scaling it for the experimental interval.

  \begin{table}%
    \centering%
    \caption{Calculations of the SI (including $Q_W$) and SD parts of the ${}^1S_0\,(F$=$1/2)\to{}^3D_1\,(F$=$1/2)$ PNC amplitude ($z$ compnent, $F_z$=0) for $^{171}$Yb ($I$=1/2) and comparison with other works. The signs have been omitted.
Units: $10^{-11}\,iea_0$}
\begin{ruledtabular}%
  \begin{tabular}{cdddl}
\multicolumn{1}{c}{} & \multicolumn{2}{c}{This work}& \multicolumn{2}{c}{Others}\\
\cline{2-3}\cline{4-5}
\multicolumn{1}{c}{} 
& \multicolumn{1}{c}{$\lambda\Sigma$\tablenote{Scaling correlation potential to reproduce the energy interval}}
& \multicolumn{1}{c}{Orthog.\tablenote{Using orthogonality conditions to subtract and re-scale the dominating term by factor $\Delta E^{\rm Calc.}/\Delta E^{\rm Exp.}$}}
& \multicolumn{1}{c}{Value}& \multicolumn{1}{c}{Ref.}\\
\hline   
\smallspace
SI-PNC &  62.5  &  59.0  &  60  	  &\cite{demille}    \\
 &    &    &  61.8  &\cite{porsev-yb-si}       \\
  &    &    &  41.6  &\cite{das-yb-si}       \\
  &    &    &  61.5  &\cite{Dzuba2011Yb}       \\
\smallspace
SD-PNC     &1.01\varkappa	&0.965\varkappa			&1.12\varkappa	&\cite{porsev-yb-sd}	\\
    &			&				&0.997\varkappa	&\cite{das-yb-sd} 	\\
    &			&				&0.990\varkappa	&\cite{Dzuba2011Yb}	\\
  \end{tabular}%
\end{ruledtabular}%
    \label{tab:compare-yb}%
  \end{table}%


To form an estimate of the uncertainty, we perform the calculations for radium, barium, and singly-ionized actinium using both the all-order and second-order correlation potentials, as described above.
The difference between these methods leads to about a 10\% difference in the PNC amplitudes.
Taking this into account, we expect the accuracy of the PNC calculations to be around 20\% for radium, barium, and singly-ionized actinium.
For thorium and protactinium, with a more complicated electron structure and less experimental data, it is harder to tell. Until a more detailed analysis can be performed these calculations should be considered order-of-magnitude estimates.
The accuracy of the calculations can be improved if measurements become available.

\section{Radium} \label{sec:ra}

Here, we study two relevant transitions for measuring PNC in radium, the $7s^2\;^1S_0 \to 6d7s\;^3D_2$ transition, and the $7s^2\;^1S_0 \to 6d7s\;^3D_1$ transition. 
Both transitions are in the optical range ($\lambda\sim700$ nm) and are enhanced by the proximity of the odd $7s7p\;^3P_1$ level to the upper levels in the transitions. 

There are several isotopes of radium that have non-zero nuclear spin. 
The nuclear spin of radium is caused by a valence neutron, which makes these transitions especially interesting for the study of the neutron-nucleus parity violating potential~\cite{AnM}. The only 
measurement of an anapole moment so far is for $^{133}$Cs, which  has only a valence proton~\cite{meas1}.

In Table~\ref{tab:ra-MEs}, we present calculations of reduced matrix elements of operators of interest for the $E1$, SI-PNC and SD-PNC interactions. 
Our calculations of the $E1$ matrix elements, as well as the energy levels (in Table~\ref{tab:energy}), agree well with previous calculations, e.g.~Ref.~\cite{BaRa-En-life,BaRa-En-E1}.

One can use the relevant values of the $E1$ and anapole moment matrix elements to determine the amplitude of the dominating term of the PNC transition. 
Note that in Table~\ref{tab:ra-MEs} we present only the electron-parts of the operators, without any additional factors. For example, the formula linking the matrix elements of $\gamma_5\rho$ to the SI-PNC interaction is
\[\bra{b}|\hat h_{SI}|\ket{a} =\frac{G_F}{2\sqrt{2}}(-Q_W)\bra{b}|\gamma_5\rho|\ket{a}.\]
We present reduced matrix elements due to their lack of dependence on the projection of angular momentum; the SD-PNC matrix elements also depend on nuclear spin. 
The reduced matrix elements obey the symmetry condition
\begin{equation*}
\bra{a}|\hat h| \ket{b} = (-1)^{J_b-J_a}\bra{b}|\hat h| \ket{a}^*,
\end{equation*}
where the asterisk stands for complex conjugation and results in a change of sign for the PNC matrix elements but not for the $E1$ matrix elements. 
Also note that the actual matrix elements contain factors depending on the different angular momentum values, for example, the SI-PNC matrix element contains the Wigner $3j$ symbol that has a term $1/\sqrt{2J+1}$, which makes these reduced matrix elements appear larger for large values of $J$.
Full formulas are given in the appendix.

The dominating contribution to the ${}^1S_0 - {}^3D_2$ transition is given by:
\begin{align}
&E_{\rm PNC}^{F_a\to F_b} \simeq	 k_{SD}
\frac{\bra{^3D_2} | \v{\alpha}\rho|  \ket{^3P_1^o}  \bra{^3P_1^o}  |-e\v{r}|  \ket{^1S_0}}
			{E(^3D_2)-E(^3P_1^o)},
\label{eq:main}
\end{align} 
where $k_{SD}$ is the coefficient (for the $z$ component):
\begin{align}
k_{\rm SD} =&
	\frac{G_F}{\sqrt{2}}\varkappa
  	\sqrt{(I+1)(2I+1)(2F_b+1)(2F_a+1)/I}
	\notag\\
&\times (-1)^{F_b - F_z} \threej{F_b}{1}{F_a}{-F_z}{0}{F_z} \notag\\
& \times
	 (-1)^{F_b-F_a}  
	\sixj{J_n}{J_b}{1}{I}{I}{F_b}
	\sixj{J_n}{J_a}{1}{F_a}{F_b}{I}
\label{eq:maincoef}
\end{align} 
(see equations (\ref{eq:pnc-rme}) and (\ref{eq:sd-pnc-rme}) in the appendix). Here $F_z={\rm min}(F_a,F_b)$, and the index $n$ refers to the intermediate state.

 Due to the large dominating term in the transitions in radium this gives a good first approximation, and was the method used in the earlier works~\cite{FlambaumRa1999,DzubaRa2000}.
We however proceed to calculate the entire sum, using the rescaled correlation potential method to deal with the numerical sensitivity due to the close opposite parity levels, as discussed above.
Table~\ref{tab:ra-test-method} compares the different methods of determining the dominating contribution to the amplitude as described above. 
There is very good agreement between these approaches, indicating good numerical accuracy in the calculations.

  \begin{table}%
    \centering%
    \caption{Reduced matrix elements $\bra{a}|\hat H|\ket{b}$ for the 
amplitudes between the lowest few states of Ra. No value means forbidden by selection rules. (a.u.)} 
\begin{ruledtabular}%
  \begin{tabular}{rlllddd}
\multicolumn{2}{c}{Even state}  & \multicolumn{2}{c}{Odd state} & \multicolumn{3}{c}{$ H_{ab}$}  \\
\cline{5-7}
\multicolumn{2}{c}{$a$}  & \multicolumn{2}{c}{$b$} & \multicolumn{1}{c}{$-e\v{r}$\tablenote{For $E1$ transition.}} & \multicolumn{1}{c}{$\gamma_5\rho$\tablenote{For SI-PNC contribution.}} & \multicolumn{1}{c}{$\v{\alpha}\rho$\tablenote{For anapole moment contribution.}} \\
\hline   
\smallspace
$7s^2$  &  $^1S_0$  &  $7s7p$  &  $^3P_0$  &    &  -22.8  &  	\\
  &    &    &  $^3P_1$  &  1.22  &    &  46.4	\\
  &    &    &  $^1P_1$  &  -5.49  &    &  -12.3	\\
\smallspace
$7s6d$  &  $^3D_1$  &  $7s7p$  &  $^3P_0$  &  2.99  &    &  3.36	\\
  &    &    &  $^3P_1$  &  -2.57  &  4.13  &  -5.45	\\
  &    &    &  $^3P_2$  &  -0.688  &    &  -1.77	\\
  &    &    &  $^1P_1$  &  -0.440  &  -9.15  &  4.74	\\
\smallspace
$7s6d$  &  $^3D_2$  &  $7s7p$  &  $^3P_1$  &  4.38  &    &  -2.21	\\
  &    &    &  $^3P_2$  &  2.60  &  -4.31  &  0.656	\\
  &    &    &  $^1P_1$  &  0.797  &    &  6.44	\\
\smallspace
$7s6d$  &  $^3D_3$  &  $7s7p$  &  $^3P_2$  &  -6.35  &    &  8.11	\\
\smallspace
$7s6d$  &  $^1D_2$  &  $7s7p$  &  $^3P_1$  &  -0.353  &    &  -4.68	\\
  &    &    &  $^3P_2$  &  -0.519  &  -5.24  &  -5.24	\\
  &    &    &  $^1P_1$  &  -3.23  &    &  -13.5	\\
  \end{tabular}%
\end{ruledtabular}%
    \label{tab:ra-MEs}%
  \end{table}%

  \begin{table}%
    \caption{Comparison of the different methods of determining the dominating term of the PNC amplitudes for $^{225}$Ra. } 
\begin{ruledtabular}%
  \begin{tabular}{cddd}
  \multicolumn{1}{c}{SD-PNC} &   \multicolumn{3}{c}{units: $10^{-10}\varkappa\, iea_0$}	\\ 
\hline
 &   \multicolumn{2}{c}{Orthog.\tablenote{Subtracting the dominating term using orthogonality conditions}} &	\\ 
\cline{2-3}
  \multicolumn{1}{c}{} 
  &   \multicolumn{1}{c}{$\lambda\Sigma$\tablenote{Scaling the correlation potential to reproduce energy interval}} 
  &   \multicolumn{1}{c}{$\Sigma$\tablenote{Re-scaling dominating term by $\Delta E^{\rm Calc.}/\Delta E^{\rm Exp.}$}}
  &   \multicolumn{1}{c}{Direct\tablenote{Calculate dominating term directly using Eq.~(\ref{eq:main}) 
		}}	
\\ 
\smallspace
$^1S_{0\,F=0.5}\to {}^3D_{2\,F=1.5}$  & 5.708 &	5.873 & 	5.706	\\
$^1S_{0\,F=0.5}\to {}^3D_{1\,F=1.5}$   & 0.1551&	0.1557&	0.1551	\\
~\\
\hline
\hline\\
  \multicolumn{1}{c}{SI-PNC} &   \multicolumn{3}{c}{units: $10^{-10}(-Q_W/N) iea_0$}	\\ 
\hline
 &   \multicolumn{2}{c}{Orthog.} &	\\ 
\cline{2-3}
  \multicolumn{1}{c}{} 
&   \multicolumn{1}{c}{$\lambda\Sigma$} 
&   \multicolumn{1}{c}{$\Sigma$}
&   \multicolumn{1}{c}{Direct}	
\\ 
\smallspace
$^1S_{0}\to {}^3D_{1}$  &  13.67  &  14.00  &  13.75	\\
  \end{tabular}%
\end{ruledtabular}%
    \label{tab:ra-test-method}%
  \end{table}%
%


The $^1S_0-{}^3D_2$ transition is of particular interest for the potential measurement of the anapole moment.
It is enhanced by  very close levels of opposite parity, the interval between the upper $^3D_2$ state and the odd-parity $^3P_1$ state is just 5.41 cm$^{-1}$, and there is no SI contribution.
This leads to a huge enhancement in the parity violating signal caused by the interaction of the valence electrons with the anapole moment of the nucleus.

Calculations of the PNC amplitudes between the different hyperfine components for this transition are presented in Table~\ref{tab:ra-1s0-3d2}.
We have performed the calculation of the entire sum, as opposed to just the leading term as was done in~\cite{DzubaRa2000}.
This amplitude is very large indeed, several orders of magnitude larger than the SD amplitudes in cesium.

  \begin{table}%
    \caption{SD-PNC amplitudes ($z$ components) for the $^1S_0\to{}^3D_2$ transition in radium, with $F_z$=min($F_a$,$F_b$). There is no SI contribution to this transition. Units: $10^{-10}iea_0\,\varkappa$} 
\begin{ruledtabular}%
  \begin{tabular}{ddddd}
  \multicolumn{1}{c}{$I$}  &    \multicolumn{1}{c}{$F_a$}  &    \multicolumn{1}{c}{$F_b$}  &    \multicolumn{1}{c}{This work}  &  \multicolumn{1}{c}{Ref.~\cite{DzubaRa2000}} \\
\hline
\smallspace
1.5  &  1.5  &  0.5  &  -1.39  &  -0.57  \\
  &    &  1.5  &  -3.35  &  -1.37  \\
  &    &  2.5  &  3.13  &  1.28  \\
\smallspace
0.5  &  0.5  &  1.5  &  5.92  &  2.42  \\
  \end{tabular}%
\end{ruledtabular}%
    \label{tab:ra-1s0-3d2}%
  \end{table}%

Our value is about twice as large as the value calculated in Ref.~\cite{DzubaRa2000}. We believe this is mainly due to the effect of the basis used for the wavefunctions on the matrix elements of the SD-PNC interaction. 
A minimal number of single-electron basis states calculated in a $V^N$ potential were used in Ref.~\cite{DzubaRa2000}. 
The use of the $V^N$ approximation in Ref.~\cite{DzubaRa2000} allowed us to have reasonable accuracy for the wave functions without saturating the basis. 
In the present work we use a complete set of single-electron states calculated in the $V^{N-2}$ potential. 
In this case, the single-electron orbitals are initially quite
different from those in the neutral atom.
However, high accuracy of the results is achieved when the basis is saturated by allowing all single and double excitations from the initial reference configuration. 
The best correspondence between two methods is achieved when only single excitations are allowed in the $V^{N-2}$ potential. 
Single excitations correct the orbitals calculated in the $V^{N-2}$ potential making them close to those calculated in the $V^N$ potential.  
In Table~\ref{tab:anm-basis} we present calculations of the $\v{\alpha}\rho$ reduced matrix element, which is proportional to the anapole moment contribution, using several different basis configurations and demonstrate that by using wavefunctions similar to those used in \cite{DzubaRa2000} we 
can account for the difference between the value determined in this work and that of Ref.~\cite{DzubaRa2000}.
Note that this change in the wavefunctions makes only a much smaller difference to the energy levels ($\sim5\%$) and $E1$ matrix elements ($\sim10\%$).

  \begin{table}%
    \caption{Effect of the basis on the matrix element $\bra{^3P_1}|\v{\alpha}\rho|\ket{^3D_2}$ in radium and comparison with Ref.~\cite{DzubaRa2000}.} 
\begin{ruledtabular}%
  \begin{tabular}{llllll}
Method
& A\tablenote{Allowing only single excitations from the main reference configuration: $7s^2$  for $^3D_2$ and $7s7p$ for $^3P_1$}
& B\tablenote{Allowing only single excitations from two reference states: $7s^2$ and $6d7s$ for $^3D_2$, and $7s7p$ and $6d7p$ for $^3P_1$}
& C\tablenote{Allowing double excitations, but with a reduced basis}
& Full\tablenote{Allowing double excitations with the full basis (final value)}
& Ref.~\cite{DzubaRa2000}  \\
\smallspace
$\bra{a}|\v{\alpha}\rho|\ket{b}$&0.90 & 1.18 & 1.48 & 2.21 & 1.10  \\
  \end{tabular}%
\end{ruledtabular}
    \label{tab:anm-basis}%
  \end{table}%

The measurement of the anapole moment of $^{133}$Cs was achieved by comparing measurements of the PNC amplitude, which contained contributions from both the SD and SI parts, between different hyperfine components~\cite{meas1}.
This transition in radium however offers the possibility to measure the effect of the anapole moment directly, which may be more efficient due both to the the much larger amplitude and to the fact that the the spin-independent interaction does not contribute in this case at all due to the large change in total electron angular momentum $\Delta J=2$.


As well as the $^1S_0-{}^3D_2$ transition, which has no SI contribution, we have also performed calculations for the $^1S_0-{}^3D_1$ transition, for which both SI and SD contributions are non-zero.
This transition is also enhanced due to close levels of opposite parity, though not to the same extent.
The interval between the even $^3D_1$ state and the odd $^3P_1$ states is 283.53 cm$^{-1}$.

These amplitudes are presented in Table~\ref{tab:ra-1s0-3d1}.
We express the amplitudes in the form
\begin{equation}
E_{\rm PNC} = P(1+R),
\end{equation}
where $P$ is the SI-PNC amplitude (including $Q_W$), and $R$ is the ratio of the SD to SI parts.
Here we calculate both parts concurrently, using the same method and wavefunctions. 
This approach has the advantage that the relative sign difference between the SI and SD parts is fixed, ensuring no ambiguity in the sign of  $\varkappa$~\cite{Dzuba2011Yb}.
There is also typically a significant improvement in accuracy for the ratio over that for each of the amplitudes individually, due to the fact that the atomic calculations for both components are very similar and much of the theoretical uncertainty cancels in the ratio~\cite{Yb+}.

The $z$ component ($J_z$=0) of the $F$-independent electron part of the spin-independent PNC amplitude for the $^1S_0-{}^3D_1$ transition in $^{223}$Ra is:
\begin{equation}
E_{\rm PNC}(^{223}{\rm Ra}) =  12.4 \times 10^{-10} (-Q_W/N)\, iea_0,   
\end{equation}
an order of magnitude larger than the $7s-8s$ transitions in Fr and Ra$^+$, and about twice as large as the $7s-6d_{3/2}$ transitions in Fr and Ra$^+$ (see, e.g.~\cite{FrLike}) and the ${}^1S_0-{}^3D_1$ transition in Yb  (see, e.g.~\cite{Dzuba2011Yb}).

  \begin{table}%
    \caption{PNC amplitudes ($z$ components) for the $^1S_0\to{}^3D_1$ transition in radium. Units: $10^{-10} iea_0$} 
\begin{ruledtabular}%
  \begin{tabular}{lldddr}
	&	  \multicolumn{1}{c}{$Q_W$}	&	  \multicolumn{1}{c}{$I$}	&	  \multicolumn{1}{c}{$F_a$}	&	  \multicolumn{1}{c}{$F_b$}	&	  \multicolumn{1}{c}{$E_{\rm PNC}$}		 \\
\hline
\smallspace
$^{223}$Ra	&	$-127.0$ 	&	1.5	&	1.5	&	0.5	&	$-6.71	\times[1-0.0402\varkappa] $\\
	&		&		&		&	1.5	&	$-9.00	\times[1-0.0161\varkappa] $\\
	&		&		&		&	2.5	&	$7.35	\times[1+0.0241\varkappa]$ \\
\smallspace
$^{225}$Ra	&	$-129.0$	&	0.5	&	0.5	&	0.5	&	$-6.81	\times[1-0.0475\varkappa]$ \\
	&		&		&		&	1.5	&	$9.64	\times[1+0.0237\varkappa]$ \\
  \end{tabular}%
\end{ruledtabular}%
    \label{tab:ra-1s0-3d1}%
  \end{table}%

\section{Barium} \label{sec:ba}

Barium, like radium, has two valence electrons above a closed shell core, and we proceed with the calculations in the same way.
Calculations of the reduced matrix elements of interest to PNC studies are presented in Table~\ref{tab:ba-MEs} (energies for Ba are presented in Table~\ref{tab:energy}). The energies and $E1$ transition amplitudes agree reasonably with previous calculations, e.g.~\cite{BaRa-En-E1}.

  \begin{table}%
    \centering%
    \caption{Reduced matrix elements $\bra{a}|\hat H|\ket{b}$ for the 
amplitudes between the relevant states of Ba. No value means forbidden by selection rules. (a.u.)} 
\begin{ruledtabular}%
  \begin{tabular}{rlllddd}
\multicolumn{2}{c}{Even state}  & \multicolumn{2}{c}{Odd state} & \multicolumn{3}{c}{$ H_{ab}$}  \\
\cline{5-7}
\multicolumn{2}{c}{$a$}  & \multicolumn{2}{c}{$b$} & \multicolumn{1}{c}{$-e\v{r}$} & \multicolumn{1}{c}{$\gamma_5\rho$} & \multicolumn{1}{c}{$\v{\alpha}\rho$} \\
\hline   
\smallspace
$6s^2$   &  ${}^1S_0$  &  $6s6p$  &  ${}^3P_0$  &    &  2.02  &  	\\
   &     &     &  ${}^3P_1$  &  -0.510  &    &  -4.12	\\
  &    &    &  ${}^1P_1$  &  5.50  &    &  1.62	\\
\smallspace
$5d6s$   &  ${}^3D_1$  &  $6s6p$  &  ${}^3P_0$  &  -2.34  &    &  -0.376	\\
  &    &     &  ${}^3P_1$  &  2.03  &  -0.245  &  0.675	\\
  &    &     &  ${}^3P_2$  &  0.532  &    &  0.335	\\
   &     &    &  ${}^1P_1$  &  0.081  &  1.07  &  -0.309	\\
  &    &  $5d6p$   &  ${}^3F_2$  &  4.23  &    &  3.89	\\
  &    &  		       &  ${}^1D_2$  &  -2.72  &    &  -0.147	\\
\smallspace
$5d6s$   &  ${}^3D_2$  &  $6s6p$  &  ${}^3P_1$  &  -3.48  &    &  0.171	\\
  &    &     &  ${}^3P_2$  &  -2.03  &  0.291  &  -0.494	\\
   &    &    &  ${}^1P_1$  &  -0.461  &    &  -0.518	\\
  &    &  $5d6p$   &  ${}^3F_2$  &  2.88  & 2.61   &  3.80	\\
  &    &                &  ${}^1D_2$  &  0.372  &  0.675  &  -0.811	\\
  &    &  		   &  ${}^3F_3$  &  -5.97  &    &  -1.76	\\
\smallspace
$5d^2$  &  ${}^1D_2$  &  $6s6p$  &  ${}^3P_1$  &  1.26  &    &  0.815	\\
  &    &     &  ${}^3P_2$  &  -1.62  &  1.29  &  1.55	\\
  &    &   &  ${}^1P_1$  &  -2.58  &    &  1.08	\\
  &    &  $5d6p$   &  ${}^3F_2$  &  -0.899  & -0.020   &  -0.427	\\
  &    &                &  ${}^1D_2$  &  -2.49  &  -0.085  &  -0.356	\\
  &    &  		   &  ${}^3F_3$  &  0.061  &    &  -0.165	\\
  \end{tabular}%
\end{ruledtabular}%
    \label{tab:ba-MEs}%
  \end{table}%

There are two transitions of interest in barium. 
The first is between the meta-stable $5d6s\,{}^3D_1$ and the upper $5d^2\,{}^1D_2$ even states.
Both SI and SD-PNC parts of this amplitude are enhanced by the proximity of the odd $5d6p\,{}^1D_2^o$ state to the upper state of the transition, with an interval of 12.34 cm$^{-1}$.

Calculations of the SI and SD contributions to the ${}^3D_1-{}^1D_2$ PNC amplitude for barium are presented in Table~\ref{tab:ba-3d1-3d2}. We present these amplitudes in the form $E_{\rm PNC}=P(1+R)$, as described above.
The $z$ component ($J_z$=1) of the $F$-independent SI-PNC amplitude for this transition is:
\begin{equation}
E_{\rm PNC}(^{135}{\rm Ba}) =  -3.55 \times 10^{-10} (-Q_W/N)\, iea_0,   
\end{equation}
despite some suppression due to the small value of the SI-PNC matrix element, it is still large.

  \begin{table}%
    \caption{PNC amplitudes ($z$ components) for the ${}^3D_1\to{}^1D_2$ transition in barium. 
			Units: $10^{-10} iea_0$} 
\begin{ruledtabular}%
  \begin{tabular}{lldddr}
	&	  \multicolumn{1}{c}{$Q_W$}	&	  \multicolumn{1}{c}{$I$}	&	  \multicolumn{1}{c}{$F_a$}	&	  \multicolumn{1}{c}{$F_b$}	&	  \multicolumn{1}{c}{$E_{\rm PNC}$}		 \\
\hline
\smallspace
$^{135}$Ba  &  $-74.0$  &  1.5  &  0.5  &  0.5  &  $2.48\times[1+0.144\varkappa]\phantom{4}$	\\
  &    &    &    &  1.5  &  $-2.48\times[1+0.0945\varkappa]$	\\
\smallspace
  &    &    &  1.5  &  0.5  &  $1.11\times[1+0.1476\varkappa]$	\\
  &    &    &    &  1.5  &  $2.66\times[1+0.0977\varkappa]$	\\
  &    &    &    &  2.5  &  $-2.49\times[1+0.0145\varkappa]$	\\
\smallspace
  &    &    &  2.5  &  1.5  &  $0.543\times[1+0.1031\varkappa]$	\\
  &    &    &    &  2.5  &  $2.18\times[1+0.0199\varkappa]$	\\
  &    &    &    &  3.5  &  $-2.51\times[1-0.0966\varkappa]$	\\
  \end{tabular}%
\end{ruledtabular}%
    \label{tab:ba-3d1-3d2}%
  \end{table}%


The other transition is from the meta-stable $5d6s\,{}^3D_2$ state to the same upper $5d^2\,{}^1D_2$ even state.
As with the first transition, both the SI and SD parts of this amplitude are enhanced by the proximity of the odd $5d6p\,{}^1D_2^o$ state to the upper state of the transition, with the same 12.34 cm$^{-1}$ interval.
However, in this case there is a second dominating term that contributes to the SD-PNC amplitude only. 
This contribution comes from the odd $5d6p\,{}^3F_3$ state, and is enhanced by an energy interval of 114.6 cm$^{-1}$.

The fact that there are two dominating terms to this transition makes this case potentially more difficult numerically -- the experimental energy intervals for both leading terms cannot be simultaneously reproduced with the same set of scaling parameters for the correlation potential. 
We proceed in this case using a mixture of the two above described methods; we use the same correlation potential scaling parameters as for the ${}^3D_1-{}^1D_2$ transition (reproducing the 12.34 cm$^{-1}$ interval exactly), and then enforce orthogonality conditions to separate off the remaining dominating term and rescale it for the 114.6 cm$^{-1}$ experimental interval.
As a test of the numerical accuracy we in fact separate off both dominating terms using the orthogonality conditions and compare them to the values calculated directly using the matrix elements from Table~\ref{tab:ba-MEs} and the experimental intervals. We find excellent agreement between these values, within 0.3\% for the SI part and better than 0.1\% for the SD part, and conclude that the numerical accuracy is good.
Calculations of the ${}^3D_2-{}^1D_2$ PNC amplitude for ${}^{135}$Ba are presented in Table~\ref{tab:ba-3d2-3d2}, and the $z$ component ($J_z$=2) of the SI-PNC amplitude for the $^3D_2-{}^1D_2$ transition for barium is:
\begin{equation}
E_{\rm PNC}(^{135}{\rm Ba}) = -0.497 \times 10^{-10} (-Q_W/N)\, iea_0,   
\end{equation}
relatively small compared to the other transitions studied in this work.

  \begin{table}%
    \caption{PNC amplitudes ($z$ components) for the ${}^3D_2\to{}^1D_2$ transition in barium. Units: $10^{-10} iea_0$} 
\begin{ruledtabular}%
  \begin{tabular}{lldddl}
	&	  \multicolumn{1}{c}{$Q_W$}	&	  \multicolumn{1}{c}{$I$}	&	  \multicolumn{1}{c}{$F_a$}	&	  \multicolumn{1}{c}{$F_b$}	&	  \multicolumn{1}{c}{$E_{\rm PNC}$}		 \\
\hline
\smallspace
$^{135}$Ba&$-74.0$
			&   1.5	&   0.5	&   0.5	&   $-0.233\times[1+0.161\varkappa]$ \\
	&   	&   	&   	&   1.5	&   $-0.233\times[1+0.190\varkappa]$  \\
\smallspace
	&   	&   	&   1.5	&   0.5	&   $-0.233\times[1+0.144\varkappa]$  \\
	&   	&   	&   	&   1.5	&   $-0.279\times[1+0.0753\varkappa]$  \\
	&   	&   	&   	&   2.5	&   $-0.174\times[1+0.146\varkappa]$  \\
\smallspace
	&   	&   	&   2.5	&   1.5	&   $-0.174\times[1+0.0686\varkappa]$  \\
	&   	&   	&   	&   2.5	&   $-0.366\times[1-0.0179\varkappa]$  \\
	&   	&   	&   	&   3.5	&   $-0.115\times[1+0.0724\varkappa]$  \\
\smallspace
	&   	&   	&   3.5	&   2.5	&  $-0.115\times[1-0.0360\varkappa]$  \\
	&   	&   	&   	&   3.5	&   $-0.466\times[1-0.131\varkappa]$  \\
  \end{tabular}%
\end{ruledtabular}%
    \label{tab:ba-3d2-3d2}%
  \end{table}%


While there are possible PNC transitions from the ground state, they are not enhanced.
The ${}^3D_1$ state is practically stable with no allowed $E1$ or $E2$ transitions to the ground state, the only lower state. We calculate the lifetime of this state to be $4\times10^{6}$ s. 
For the ${}^3D_2$ state, which has only a significant contribution from an $E2$ transition to the ground state, we calculate a lifetime of 70 s, which agrees very well with the value of 69 s calculated in Ref.~\cite{BaRa-En-life}.

Though the amplitudes are smaller than those in radium, and despite the fact that the enhanced amplitudes are not  from the ground state, there may may be advantages to working with barium.
The ${}^{135}$Ba and ${}^{137}$Ba  nuclei, each with nuclear spin $I=3/2$, are stable. 
There are obvious benefits to this over working with radioactive elements.
Also, despite the smaller amplitudes and likely smaller nuclear enhancement (i.e.~smaller $\varkappa$) than in radium, the ratio of the SD to SI parts is very large for transitions between some hyperfine components. 
This is due to small values for the SI-PNC matrix element of the dominating terms (see Table~\ref{tab:ba-MEs}), which suppresses the SI part of the amplitudes.

As well as searching for the anapole moment by measuring PNC between different hyperfine components, each of the spin-independent transitions in barium could be used for measurements of PNC in a chain of isotopes
There is currently much interest in this type of measurement, in particular for  ytterbium~\cite{Dzuba2011Yb,demille,porsev-yb-si,das-yb-si,porsev-yb-sd,das-yb-sd}, where measurements have already been performed~\cite{Yb-exp}. The atomic PNC amplitude in Yb constitutes the largest yet observed in any system.
The SI-PNC transitions in neutral barium are of particular interest in this area.
Like ytterbium, barium has many  stable isotopes, with both even and odd-nucleon numbers, that are significantly spread out.
The ${}^3D_1-{}^1D_2$ SI-PNC amplitude in barium is about half the size of the ${}^1S_0-{}^3D_1$ transition of ytterbium, though the SD contribution for barium is several times larger than that for ytterbium.

\section{Actinium II}\label{sec:ac}

Singly ionized actinium has a ground state configuration very similar to that of radium, and thus the calculations can be approached in the same way.
Here, the transition of interest is the $7s^2\,^1S_0\to6d^2\,^3P_2$ transition, for which there is no SI contribution.
This transition is enhanced by the proximity of the odd $7s7p\,^3P_1$ level to the even $6d^2\,^3P_2$ level, with an energy interval of 18.93 cm$^{-1}$.

\begin{table}%
    \centering%
    \caption{Reduced matrix elements $\bra{b}|\hat H|\ket{a}$ of the amplitudes between the lowest few states of Ac$^+$.   No value means forbidden by selection rules. (a.u.)} 
\begin{ruledtabular}%
  \begin{tabular}{rlllddd}
\multicolumn{2}{c}{Odd state}  & \multicolumn{2}{c}{Even state} & \multicolumn{3}{c}{$ H_{ba}$}  \\
\cline{5-7}
\multicolumn{2}{c}{$a$}  & \multicolumn{2}{c}{$b$} & \multicolumn{1}{c}{$-e\v{r}$} & \multicolumn{1}{c}{$\gamma_5\rho$} & \multicolumn{1}{c}{$\v{\alpha}\rho$} \\
\hline 
\smallspace
$7s7p$  &  $^3P_0$  &  $7s^2$  &  $^1S_0$  &    &  45  &    \\
  &    &  $6d7s$  &  $^3D_1$  &  -1.8  &    &  5.6  \\
  &    &  $6d^2$  &  $^3P_0$  &    &  5.2  &    \\
  &    &    &  $^3P_1$  &  -0.50  &    &  0.029  \\
\smallspace
$7s7p$  & $^3P_1$  &  $7s^2$  &  $^1S_0$  &  1.7  &    &  -85  \\
  &    &  $6d7s $ &  $^3D_1$  &  1.7  &  16  &  -15  \\
  &    &    &  $^3D_2$  &  2.2  &    &  14  \\
  &    &    &  $^1D_2$  &  -0.10  &    &  3.1  \\
  &    &  $6d^2$  & $^3F_2$  &  0.56  &    &  -3.3  \\
  &    &    &  $^3P_0$  &  0.66  &    &  12  \\
  &    &    &  $^3P_1$  &  -0.42  &  1.6  &  -1.5  \\
  &    &    &  $^3P_2$  &  0.40  &    &  -4.2  \\
  &    &    &  $^1D_2$  &  0.82  &    &  -4.7 
  \end{tabular}%
\end{ruledtabular}%
    \label{tab:acMEs}%
  \end{table}%

In Table~\ref{tab:acMEs} we present calculations of the relevant reduced matrix elements for Ac$^+$, and in Table~\ref{tab:ac-1s0-3p2} we present $z$ components of the SD-PNC amplitudes.
These amplitudes are almost as large as the corresponding ${}^1S_0-{}^3D_2$ transitions in radium.
Despite the difficulties of working with ions, it is possible that there are advantages in using actinium. 
The ${}^{227}$Ac nucleus has a half life of $22$ yr, much longer than the 42 min half-life of the most stable odd-nucleon isotope of radium ${}^{227}$Ra or the 22 min of the most stable francium isotope ${}^{223}$Fr.

This transition in Ac$^+$ could be measured using a similar method to that put forward by Fortson in Ref.~\cite{Fortson} for measuring PNC in single ions that have been laser trapped and cooled.
The upper $6d^2$ level of the transition should be relatively stable, a condition for accuracy in this method, since the only allowed $E1$ transition to a lower level is suppressed by the small interval 18.93 cm$^{-1}$.
Including all $E1$, $E2$ and $M1$ transitions to lower levels, we calculate the lifetime of this state to be about 0.2 s.

  \begin{table}%
    \caption{SD-PNC amplitudes ($z$ components) for the $^1S_0\to{}^3P_2$ transition in Ac$^+$. There is no SI contribution to this transition. Units: $10^{-10} iea_0\,\varkappa$} 
\begin{ruledtabular}%
  \begin{tabular}{ldddd}
  &  \multicolumn{1}{c}{$I$}  &  \multicolumn{1}{c}{$F_a$}  &  \multicolumn{1}{c}{$F_b$}  &   \multicolumn{1}{c}{$E_{\rm PNC}$} \\
\hline
\smallspace
$^{227}$Ac$^+$  &  1.5  &  1.5  &  0.5  &  1.05 \\
  &    &   &  1.5  &  2.51 \\
  &    &    &  2.5  &  -2.35 \\
\smallspace
$^{231}$Ac$^+$  &  0.5  &  0.5  &  1.5  &  -4.44 \\
  \end{tabular}%
\end{ruledtabular}%
    \label{tab:ac-1s0-3p2}%
  \end{table}%

\section{Thorium} \label{sec:th}

Thorium has four valence electrons.
Full-scale  accurate calculations for this atom are beyond the scope of the present work, however we use the same methods outlined above (using a $V^{N-4}$ potential) to perform preliminary calculations  here also.

To perform the calculations of the PNC amplitudes we calculate only the dominating contribution using the matrix elements of the PNC and $E1$ interactions, without trying to evaluate the entire sum.
For the wavefunctions we include the eight leading configurations and from these allow single excitations. This provides a fair compromise between completeness of the wavefunctions and ease of computation.
Here we include correlation corrections, but we calculate these to second order in MBPT only. 
We also do not perform any re-scaling of the correlation potential.
This is because the uncertainty here is dominated by the completeness of the basis, not by the effect of the correlation potential.
We have performed calculations of several energy levels of interest to PNC in thorium. 
These are presented in Table~\ref{tab:th-energy} (note that this is not a comprehensive list of states). 
Despite the lower level approximation for the more complex system the agreement is reasonably good, particularly for the lower states.
Calculations of the relevant $E1$ and PNC reduced matrix elements are presented in Table~\ref{tab:th-MEs}.

It is worth noting that for thorium (and also for protactinium) the configuration mixing is very large, particularly for the higher states. 
The configurations given in Tables~\ref{tab:th-energy} and \ref{tab:th-MEs} are the leading configurations (taken from Ref.~\cite{BWParis}), but other contributing configurations are important as well.
For example, the SI-PNC ($\gamma_5\rho$) reduced matrix element between the $6d7s^27p$    $^3D_3$ and $6d^27s^2$ $^3F_3$ states is rather large (see Table~\ref{tab:th-MEs}), which is unexpected since the leading configurations suggest this transition is essentially a single-electron $p$-$d$ transition. 
(This particular matrix element does not contribute to the PNC amplitude we study here.)
However, due to the large configuration mixing this matrix element also has a large contribution coming from single-electron $s$-$p$ mixing, enhancing this amplitude. 
The extent of this mixing is detrimental to the accuracy of the calculations, especially when the overall accuracy is not high, since even relatively small errors in the calculated configurations may lead to large errors in the weak matrix elements (as discussed in Section~\ref{sec:theory}). 
 Fortunately, the weak matrix elements of relevance to the PNC amplitudes studied here are relatively stable in this regard. Still, this is a large contributing factor to the low accuracy for these calculations in thorium and protactinium.

  \begin{table}%
    \centering%
    \caption{Calculated energy levels for thorium and comparison with experiment (Ref.~\cite{BWParis}). 
Units are cm$^{-1}$.} 
\begin{ruledtabular}%
  \begin{tabular}{llrr}
 \multicolumn{2}{c}{State}  & \multicolumn{1}{c}{Calc.} & \multicolumn{1}{c}{Exp.} \\
\hline   
\smallspace
$6d^27s^2$  &  $^3F_2$  &  0  &  0  \\
&	$^3P_0$	& 2546	&2558 \\
  &  $^3F_3$  &  3168  &  2869  \\
 &	$^1P_2$
				&	4120 &	3688 \\
 &	$^3P_1$	& 3926	 &3865 \\
  &  $^3F_4$  &  5650  &  4962  \\
\smallspace
$6d^37s$	&$^5F_1$& 5257	&		5563 \\
&	$^5F_2$&	6232&	6362 \\
 &  $^3H_4$  &  18358  &  15493  \\
  &  $^3F_3$  &  18536  &  17398  \\
\smallspace
$5f6d7s^2$  &  $^3G_5^o$  &  18846  &  15490  \\
\smallspace
$6d7s^27p$  &  $^3D_3^o$  &  20190  &  17411  \\
  \end{tabular}%
\end{ruledtabular}%
    \label{tab:th-energy}%
  \end{table}%

\begin{table}%
    \centering%
    \caption{Reduced matrix elements $\bra{b}|\hat H|\ket{a}$ of the amplitudes between the relevant states of Th. (a.u.)} 
\begin{ruledtabular}%
  \begin{tabular}{rlllddd}
\multicolumn{2}{c}{Odd state}  & \multicolumn{2}{c}{Even state} & \multicolumn{3}{c}{$ H_{ba}$}  \\
\cline{5-7}
\multicolumn{2}{c}{$a$}  & \multicolumn{2}{c}{$b$} & \multicolumn{1}{c}{$-e\v{r}$} & \multicolumn{1}{c}{$\gamma_5\rho$} & \multicolumn{1}{c}{$\v{\alpha}\rho$} \\
\hline 
\smallspace
$6d7s^27p$  &  $^3D_3$  &  $6d^27s^2$  &  $^3F_2$  &  -0.17  &    &  -14  \\
  &    &    &  $^3F_3$  &  -0.20  &  -47  &  2.8  \\
  &    &    &  $^3F_4$  &  -0.51  &    &  -32  \\
  &    &  $6d^37s$  &  $^3H_4$  &  -0.01  &    &  -7.1  \\
  &    &    &  $^3F_3$  &  0.15  &  -0.37  &  4.5  \\
\smallspace
$5f6d7s^2$  &  $^3G_5$  &  $6d^27s^2$  &  $^3F_4$  &  1.7  &    &  53  \\
  &    &  $6d^37s$  &  $^3H_4$  &  0.14  &    &  2.9  \\

  \end{tabular}%
\end{ruledtabular}%
    \label{tab:th-MEs}%
  \end{table}%

Thorium has several isotopes of non-zero nuclear spin, e.g., $^{227}$Th with $I=1/2$,  $^{225}$Th with $I=3/2$, and $^{229}$Th  with $I=5/2$. 
The most long-lived of these is $^{229}$Th, which has a half-life of about $7300$ yr, and the most stable thorium isotope is $^{232}$Th, which has zero nuclear spin and a half life of 1.4$\times10^{10}$ yr.
The nuclear spin of thorium is produced by a valence neutron.
There are two interesting PNC transitions in thorium that are enhanced by close opposite parity levels.

The first is a transition between the meta-stable $6d^27s^2\,{}^3F_4$ state and the higher $6d^37s\,{}^3H_4$ state, which is induced by the nuclear anapole moment.
This transition is enhanced by the extremely small 3.1~cm$^{-1}$ energy interval between the $5f6d7s^2\,{}^3G_5^o$ state and the ${}^3H_4$ state.
This transition is given by
\begin{align}
E_{\rm PNC} &\simeq	 k_{SD}
\frac{\bra{^3H_4} | \v{\alpha}\rho|  \ket{^3G_5^o}  \bra{^3G_5^o}  |-e\v{r}|  \ket{^3F_4}}
			{E(^3H_4)-E(^3G_5^o)},      
\label{eq:main-th}
\end{align} 
where the factor $ k_{SD}$ is given in Eq.~(\ref{eq:maincoef}). 
The SD-PNC amplitudes between various hyperfine components for this transition are given in Table~\ref{tab:th-anm-sd}. 
There is also a SI contribution to this transition, though it is not enhanced by the proximity of opposite parity levels and doesn't contain a single dominating term. 
It is likely between one and two orders of magnitude smaller than the SD contribution, below the current level of accuracy.

The second transition of interest in thorium is between the ground $6d^27s^2\,{}^3F_2$ state and the higher $6d^37s\,{}^3F_3$ state. 
This transition is enhanced by a 12.8~cm$^{-1}$ energy interval between the $6d7s^27p\,{}^3D_3^o$ state and the ${}^3F_3$ state, and has contributions from both the anapole moment and the nuclear weak charge.
The amplitudes for this transition are presented in Table~\ref{tab:th-qw-anm}, where, as for radium, we present the amplitudes in the form $E_{\rm PNC} = P(1+R)$. Note that the ratio of the SD contribution to the SI contribution is significantly larger for thorium than for radium.

The $z$ component ($J_z$=2) of the SI-PNC amplitude for the ($F$-independent) ${}^3F_2 - {}^3F_3$ transition in $^{232}$Th is calculated to be
\begin{equation}
E_{\rm PNC}(^{232}{\rm Th}) =   9.9\times 10^{-11} (-Q_W/N)\, iea_0.   %
\end{equation}
There is no SD-PNC contribution here due to the fact that ${}^{232}$Th has nuclear spin $I=0$.
Despite suppression from both the $E1$ and PNC matrix elements the amplitude is large, an order of magnitude larger than the $6s-7s$ transition in cesium.
Though the accuracy here is not high, this transition could be used in isotopic chain measurements to determine ratios of the weak charges for different isotopes of thorium. 
$^{232}$Th is practically stable, with a half-life of 1.5$\times10^{10}$ yr.

  \begin{table}%
    \caption{SD-PNC amplitudes ($z$ components) for the $^3F_4\to{}^3H_4$ transition in thorium. 
The SI contribution to this transition is not enhanced. 
Units: $10^{-10} iea_0\,\varkappa$} 
\begin{ruledtabular}%
  \begin{tabular}{ldddd}
\multicolumn{1}{c}{}   &  \multicolumn{1}{c}{$I$}  &  \multicolumn{1}{c}{$F_a$}  &  \multicolumn{1}{c}{$F_b$}  &   \multicolumn{1}{c}{$E_{\rm PNC}$} \\
\hline
\smallspace
$^{227}$Th &0.5&3.5&4.5& 3.4\\
\smallspace
 & & 4.5   &  4.5  &   -1.0 \\
\smallspace
$^{229}$Th&2.5  &  1.5  &  2.5  &  1.4  \\
\smallspace
 & & 2.5   &  2.5  &   -0.87 \\
 & &        & 3.5  &  1.7  \\
\smallspace
&  &  3.5  &  2.5  &  -0.12  \\
&  &        &  3.5  &  -1.4  \\
 & &        &  4.5  &  1.8  \\
\smallspace
 & & 4.5   & 3.5   & -0.16 \\
 & &        & 4.5   &  -1.7  \\
 & &       & 5.5   &  1.8  \\
\smallspace
 & & 5.5   & 4.5   &  -0.14  \\
 & &        &   5.5 &  -1.5  \\
 & &        &   6.5 &  1.5  \\
\smallspace
 & & 6.5   &  5.5  & -0.086   \\
 & &        &  6.5  &  -1.0  \\
  \end{tabular}%
\end{ruledtabular}%
    \label{tab:th-anm-sd}%
  \end{table}%

  \begin{table}%
    \caption{PNC amplitudes ($z$ components) for the ${}^3F_2\to{}^3F_3$ transition in thorium. 
Units: $10^{-10} iea_0$} 
\begin{ruledtabular}%
  \begin{tabular}{lldddr}
	&	  \multicolumn{1}{c}{$Q_W$}	&	  \multicolumn{1}{c}{$I$}	&	  \multicolumn{1}{c}{$F_a$}	&	  \multicolumn{1}{c}{$F_b$}	&	  \multicolumn{1}{c}{$E_{\rm PNC}$}		 \\
\hline
\smallspace
$^{227}$Th	&  $-128.9$	&  0.5	&  1.5	&  2.5	&  $0.95\times[1+0.22\varkappa]$  \\
\smallspace  
	&  	&  	&  2.5	&  2.5	&  $-0.34\times[1+0.22\varkappa]$  \\
	&  	&  	&  	&  3.5	&  $0.83\times[1-0.16\varkappa]$  \\
\smallspace  
$^{229}$Th	&  $-130.8$	&  2.5	&  0.5	&  0.5	&  $-0.79\times[1+0.21\varkappa]$  \\
	&  	&  	&  	&  1.5	&  $0.72\times[1+0.18\varkappa]$  \\
\smallspace  
	&  	&  	&  1.5	&  0.5	&  $-0.42\times[1+0.21\varkappa]$  \\
	&  	&  	&  	&  1.5	&  $-0.91\times[1+0.18\varkappa]$  \\
	&  	&  	&  	&  2.5	&  $0.68\times[1+0.13\varkappa]$  \\
\smallspace  
	&  	&  	&  2.5	&  1.5	&  $-0.28\times[1+0.18\varkappa]$  \\
	&  	&  	&  	&  2.5	&  $-0.91\times[1+0.13\varkappa]$  \\
	&  	&  	&  	&  3.5	&  $0.68\times[1+0.054\varkappa]$  \\
\smallspace  
	&  	&  	&  3.5	&  2.5	&  $-0.17\times[1+0.13\varkappa]$  \\
	&  	&  	&  	&  3.5	&  $-0.83\times[1+0.053\varkappa]$  \\
	&  	&  	&  	&  4.5	&  $0.68\times[1-0.042\varkappa]$  \\
\smallspace  
	&  	&  	&  4.5	&  3.5	&  $-0.091\times[1+0.053\varkappa]$  \\
	&  	&  	&  	&  4.5	&  $-0.62\times[1-0.043\varkappa]$  \\
	&  	&  	&  	&  5.5	&  $0.68\times[1-0.16\varkappa]$  \\
  \end{tabular}%
\end{ruledtabular}%
    \label{tab:th-qw-anm}%
  \end{table}%

\section{Protactinium}\label{sec:pa}

As well as the enhancement that is due to the presence of close electronic levels of opposite parity, there is some suggestion that there may also be a very large nuclear enhancement of $P$-, $T$-odd effects in ${}^{229}$Pa~\cite{ThPa1} (see also~\cite{ThPa2,ThPa3,CloseNuc84,CloseNuc83}), which has nuclear spin $I=5/2$.
The suggestion of large nuclear enhancement comes from experimental evidence that there is an extremely small energy splitting ($\sim0.22$ keV) between the members of a ground state parity doublet~\cite{PaExp1}.
However, more recent experimental work  has put the identification of these levels into doubt (see, e.g.~\cite{PaExp2,PaExp3}).
Even so, the parity violating nuclear effects can reasonably be expected to be large, and along with the electronic enhancement this makes protactinium an interesting case also. 

To perform these calculations for protactinium, which has five valence electrons, we follow a similar procedure as in thorium, however we do not allow any excitations from the eight leading configurations in the production of the wavefunctions.
The experimental energies of these states, as well as  calculations of the reduced matrix elements of the relevant operators, are presented in Table~\ref{tab:pa-MEs}.

Note that it would be preferable to perform calculations in protactinium (and even thorium) using the conventional CI method with a $V^N$ potential.
The benefits for this type of Hartree-Fock potential when only a small basis is used for the valence wavefunctions was discussed in Section~\ref{sec:ra}.
However, we find that in these cases the convergence of the TDHF equations (\ref{eq:RPA}) is problematic due to the open $s$, $f$ and $d$ shells of the important configurations.
This is especially true for the operator of the SD-PNC interaction, which leads to unstable and unreliable results.
It is for this reason we use the $V^{N-M}$ potential approach despite the reduction in accuracy.

\begin{table}%
    \centering%
    \caption{Reduced matrix elements $\bra{a}|\hat H|\ket{b}$ for the amplitudes between the relevant states of Pa.   Also shown are the experimental energies of the levels~\cite{BWParis}.} 
\begin{ruledtabular}%
  \begin{tabular}{llllddr}
\multicolumn{2}{c}{Odd state}  & \multicolumn{2}{c}{Even state} & \multicolumn{3}{c}{$ H_{ab}$ (a.u.)}  \\
\cline{5-7}
\multicolumn{2}{c}{$a$}  & \multicolumn{2}{c}{$b$} 
& \multicolumn{1}{c}{$-e\v{r}$} 
& \multicolumn{1}{c}{$\gamma_5\rho$} 
& \multicolumn{1}{c}{$\v{\alpha}\rho$} \\
\hline 
\smallspace
$5f6d^37s$  &  ${}^6I_{9/2}^o$   &   $5f^26d7s^2$  &  $^4K_{11/2}$&  0.22   && -13.7  \\
\multicolumn{2}{c}{$(8583~{\rm cm}^{-1})$}  &    \multicolumn{2}{c}{$(0~{\rm cm}^{-1})$} &  &      &    \\
&&&&&\\
 &     &   $5f^26d7s^2$  &  $^4G_{11/2}$& 0.36  && -4.5 \\
\multicolumn{2}{c}{}  &    \multicolumn{2}{c}{$(8571~{\rm cm}^{-1})$} &      &  &   \\
&&&&&\\
 &     &   $5f^26d7s^2$  &  $^4H_{9/2}$&  -0.08  &2.8 &3.8 \\
\multicolumn{2}{c}{}  &    \multicolumn{2}{c}{$(8596~{\rm cm}^{-1})$} &        &  &\\
  \end{tabular}%
\end{ruledtabular}%
    \label{tab:pa-MEs}%
  \end{table}%

 \begin{table}%
    \caption{SD-PNC amplitudes ($z$ components) for the $^4K_{11/2}\to{}^4G_{11/2}$ transition in protactinium. 
The SI contribution to this transition is not enhanced. 
Units: $10^{-10} iea_0\,\varkappa$} 
\begin{ruledtabular}%
  \begin{tabular}{ldddd}
\multicolumn{1}{c}{}   &  \multicolumn{1}{c}{$I$}  &  \multicolumn{1}{c}{$F_a$}  &  \multicolumn{1}{c}{$F_b$}  &   \multicolumn{1}{c}{$E_{\rm PNC}$} \\
\hline
\smallspace
$^{229}$Pa &2.5  &  3  &  3  &  0.039  \\
			&       &     &  4  &  0.0061  \\
\smallspace
&  &  4  &  3  & -0.055  \\
&  &    &   4 &  0.063 \\
&  &    &  5  & 0.0070  \\
\smallspace
&  &  5  &  4  & -0.071 \\
&  &    &   5 &  0.072 \\
&  &    &  6  &  0.0056 \\
\smallspace
&  & 6  &  5  & -0.078  \\
&  &    &   6 & 0.067  \\
&  &    &  7 & 0.0028  \\
\smallspace
&  &  7  &   6 &  -0.077 \\
&  &    &  7 &  0.043 \\
\smallspace
&  &  8  &  7 & -0.064  \\
\smallspace
$^{231}$Pa&1.5  &  4  &  4  & 0.041   \\
&  &    &  5  &  0.0032  \\
\smallspace
&  &  5  &  4  &  -0.078  \\
&  &    &   5 &  0.056  \\
&  &    &  6  & 0.0024  \\
\smallspace
&  &  6  & 5   &  -0.092  \\
&  &    &   6 &  0.043  \\
  \end{tabular}%
\end{ruledtabular}%
    \label{tab:pa-anm-sd}%
  \end{table}%

  \begin{table}%
    \caption{PNC amplitudes ($z$ components) for the ${}^4K_{11/2}\to{}^4H_{9/2}$ transition in protactinium.
Units: $10^{-10} iea_0$} 
\begin{ruledtabular}%
  \begin{tabular}{lldddr}
	&	  \multicolumn{1}{c}{$Q_W$}	&	  \multicolumn{1}{c}{$I$}	&	  \multicolumn{1}{c}{$F_a$}	&	  \multicolumn{1}{c}{$F_b$}	&	  \multicolumn{1}{c}{$E_{\rm PNC}$}		 \\
\hline
\smallspace
$^{229}$Pa	&  $-129.8$	&  2.5	&  3	&  2	&  $-5.1\times[1+0.023\varkappa]$\\
	&  	&  	&  	&  3	&  $3.0\times[1+0.018\varkappa]$\\
	&  	&  	&  	&  4	&  $0.35\times[1+0.011\varkappa]$\\
\smallspace  
	&  	&  	&  4	&  3	&  $-4.4\times[1+0.018\varkappa]$\\
	&  	&  	&  	&  4	&  $3.5\times[1+0.011\varkappa]$\\
	&  	&  	&  	&  5	&  $0.32\times[1+0.0028\varkappa]$\\
\smallspace  
	&  	&  	&  5	&  4	&  $-3.8\times[1+0.011\varkappa]$\\
	&  	&  	&  	&  5	&  $3.5\times[1+0.0028\varkappa]$\\
	&  	&  	&  	&  6	&  $0.25\times[1-0.0073\varkappa]$\\
\smallspace  
	&  	&  	&  6	&  5	&  $-3.8\times[1+0.0028\varkappa]$\\
	&  	&  	&  	&  6	&  $3.0\times[1-0.0073\varkappa]$\\
	&  	&  	&  	&  7	&  $0.15\times[1-0.019\varkappa]$\\
\smallspace  
	&  	&  	&  7	&  6	&  $-3.5\times[1-0.007\varkappa]$\\
	&  	&  	&  	&  7	&  $2.3\times[1-0.019\varkappa]$\\
\smallspace  
	&  	&  	&  8	&  7	&  $-3.5\times[1-0.019\varkappa]$\\
\smallspace  
$^{231}$Pa	&  $-131.8$	&  1.5	&  4	&  3	&  $-4.7\times[1+0.023\varkappa]$\\
	&  	&  	&  	&  4	&  $2.2\times[1+0.012\varkappa]$\\
	&  	&  	&  	&  5	&  $0.14\times[1-0.0022\varkappa]$\\
\smallspace  
	&  	&  	&  5	&  4	&  $-4.1\times[1+0.012\varkappa]$\\
	&  	&  	&  	&  5	&  $2.3\times[1-0.0022\varkappa]$\\
	&  	&  	&  	&  6	&  $0.10\times[1-0.019\varkappa]$\\
\smallspace  
	&  	&  	&  6	&  5	&  $-3.8\times[1-0.002\varkappa]$\\
	&  	&  	&  	&  6	&  $1.9\times[1-0.019\varkappa]$\\
\smallspace  
	&  	&  	&  7	&  6	&  $-3.8\times[1-0.019\varkappa]$\\

  \end{tabular}%
\end{ruledtabular}%
    \label{tab:pa-qw-anm}%
  \end{table}%

There are two transitions of particular interest in protactinium.
The first is between the even $5f^26d7s^2\,{}^4K_{11/2}$ ground-state and the $5f^26d7s^2\,{}^4G_{11/2}$ upper state, and the other is between the ground-state and the $5f^26d7s^2\,{}^4H_{9/2}$ state.
Both transitions are enhanced by the proximity of the odd $5f6d^37s\,{}^6I_{9/2}^o$ state to the upper state of the transitions with energy intervals of 12.0~cm$^{-1}$ and 13.2~cm$^{-1}$ respectively.

Calculations of the anapole moment induced PNC amplitudes for the ${}^4K_{11/2}-{}^4G_{11/2}$ transition are presented in Table~\ref{tab:pa-anm-sd}. 
There is also a SI contribution to this transition but it is not enhanced and is smaller than the SD part.

The ${}^4K_{11/2}-{}^4H_{9/2}$ transition transition has enhanced contributions from both the anapole moment and $Q_W$ induced contributions. 
The amplitudes for this transition are presented in Table~\ref{tab:pa-qw-anm}.
The SD amplitudes in Table~\ref{tab:pa-anm-sd} and~\ref{tab:pa-qw-anm} are approximately 10 times smaller than those calculated for thorium, however the anapole moment, $\kappa_a({}^{229}$Pa), may be much larger.

We calculate the $z$ component ($J_z$=9/2) for the $F$-independent part of the  ${}^4K_{11/2}-{}^4H_{9/2}$ SI-PNC amplitude to be
\begin{equation}
E_{\rm PNC}(^{231}{\rm Pa}) =   -44\times 10^{-11} (-Q_W/N)\, iea_0.   %
\end{equation}
As for Th, the accuracy here is not high, though the amplitude is very large, about the same size as the $7s-6d_{3/2}$ amplitudes in Fr and Ra$^+$, and about a third of the size of the $^1S_0-{}^3D_1$ amplitude in neutral radium.
This transition would therefore be of interest for measuring the ratio weak charges for a number of different isotopes of Pa, the most stable of which being ${}^{231}$Pa, with a half-life of about 32500 yr.

\section*{Conclusion}

We have presented calculations of strongly enhanced atomic parity nonconservation due both to the nuclear weak charge and the nuclear anapole moment in the hope of motivating experiment.
Anapole moment induced transitions are presented for systems whose nuclear spin is caused  both by a valence neutron (Ra, Ba, and Th) and a proton (Ac$^+$ and Pa).
We expect the accuracy of the calculations to be approximately 20\% for Ra, Ba, and Ac$^+$, and provide order-of-magnitude calculations for Th and Pa.
Calculations for thorium and protactinium can be improved by extending the CI calculations and performing a summation for the entire PNC amplitude. 
More complete calculations for all systems can be performed if experimental work is under way.
Due to the very large enhancement of the PNC amplitudes the atoms and ions considered here are promising candidates for experimental studies of parity violating nuclear forces and for studying PNC in a chain of isotopes.

\acknowledgments
One of the  authors (V.A.D.) would like to express a special thanks to
the Mainz Institute for Theoretical Physics (MITP) for its hospitality
and support. The work was also supported by the Australian Research Council.

\appendix

\section{Formulas}

The parity violating $E1$ transition induced by the $h_{\rm PNC}$ interaction (\ref{eq:hPNC}) is given by equation~(\ref{eq:fullsummation}).
With use of the Wigner-Eckart theorem, this amplitude can be expressed via the reduced matrix elements:
\begin{equation}
E_{\rm PNC} = (-1)^{F_b - M_b} \threej{F_b}{1}{F_a}{-M_b}{q}{M_a}\bra{J_bF_b}|d_{\rm PNC}| \ket{J_aF_a}.
\label{eq:pnc-rme}
\end{equation}
For $z$ components we take $M={\rm min}(F_a,F_b).$

The reduced matrix elements obey the symmetry rule
\begin{align}
\bra{J_aF_a}|d_{\rm PNC}| \ket{J_bF_b} &= (-1)^{F_b-F_a}\bra{J_bF_b}|d_{\rm PNC}| \ket{J_aF_a}^*,
\end{align}
where ${}^*$ means complex conjugation and results in a change of sign for the PNC amplitudes.

For the single-electron wavefunctions we use the form:
\begin{equation}
\psi_{jlm}(\v{r}) =\begin{pmatrix}f(r)\Omega(\v{r}/r)_{\kappa m}\\i\alpha g(r) \Omega(\v{r}/r)_{-\kappa m}\end{pmatrix},
\end{equation}
where $\alpha\approx1/137$ is the fine-structure constant, and
$\kappa=\mp(j+1/2)$ for $j=1\pm 1/2$ is the Dirac quantum number.

\subsection{Spin-dependent PNC} 

For the SD-PNC amplitude, the reduced matrix element is given by
\begin{align}
&\bra{J_bF_b}| d_{\rm SD}  | \ket{J_aF_a} 	
	\notag\\
	&=
	\frac{G_F}{\sqrt{2}}\varkappa
  	\sqrt{(I+1)(2I+1)(2F_b+1)(2F_a+1)/I}
	\notag\\
& \times
	\sum_n \Bigg[ (-1)^{J_b-J_a}  
	\sixj{J_n}{J_a}{1}{I}{I}{F_a}
	\sixj{J_n}{J_b}{1}{F_b}{F_a}{I}
	\notag\\
		& \phantom{\times\sum_n	\Bigg[ \;} 
			\times \frac{\bra{J_b} |\hat{d}_{\rm E1}|  \ket{J_n} \bra{J_n}  |\v{\alpha}\rho|  \ket{J_a}}
			{E_{a}-E_{n}}    \notag\\
		& \phantom{\times\sum_n	\Bigg[ \;} 
		+
(-1)^{F_b-F_a}  
	\sixj{J_n}{J_b}{1}{I}{I}{F_b}
	\sixj{J_n}{J_a}{1}{F_a}{F_b}{I}
		\notag\\
		& \phantom{\times\sum_n	\Bigg[ \;}
			\times\frac{\bra{J_b} | \v{\alpha}\rho|  \ket{J_n}  \bra{J_n}  |\hat{d}_{\rm E1}|  \ket{J_a}}
			{E_{b}-E_{n}}\Bigg].
\label{eq:sd-pnc-rme}
\end{align}

The single-electron contributions to the reduced matrix element of the SD-PNC interaction has the form
\begin{equation}
\bra{J_a} | \v{\alpha}\rho|  \ket{J_b} = iR_{ 1SD} C_{ 1SD} + iR_{ 2SD} C_{ 2SD}, 
\end{equation}
where 
\begin{align}
R_{ 1SD} &= -\alpha\int \rho g_af_b \,{\rm d}r, \notag\\ 
R_{ 2SD} &= -\alpha\int \rho f_ag_b \,{\rm d}r,
\end{align}
are the radial integrals with $\rho(r)$ the  (Fermi-type) nuclear density normalized to 1,  and $C_{1,2SD}$ are the angular coefficients:
\begin{align*}
C_{ 1SD} &= (-1)^{J_a + l_b+1/2}\sqrt{6(2J_a+1)(2J_b+1)}\\&\times\sixj{1/2}{J_a}{l_b}{J_b}{1/2}{1}\\
C_{ 1SD} &= (-1)^{J_a + l_a+3/2}\sqrt{6(2J_a+1)(2J_b+1)}\\&\times\sixj{1/2}{J_a}{l_a}{J_b}{1/2}{1}.
\end{align*}

\subsection{Spin-independent PNC}

For the $Q_W$ induced SI amplitude, the reduced matrix element is given by
\begin{align}
&\bra{J_bF_b}| d_{\rm SI}  | \ket{J_aF_a} =
i\frac{G_F}{2\sqrt{2}}(-Q_W)(-1)^{I+F_a + J_b+1}  \notag\\
&\times\sqrt{(2F_b+1)(2F_a+1)}\sixj{J_a}{J_b}{1}{F_b}{F_a}{I}  
\notag\\
& \times
	\sum_n 
	\Bigg[
		\threej{J_a}{0}{J_a}{-m}{0}{m}  
			\frac{
			\bra{J_b} |\hat{d}_{\rm E1}|  \ket{J_n} \bra{J_n}  |\gamma_5\rho|  \ket{J_a}}
			{E_{a}-E_{n}}  
		\notag\\
		& \phantom{\times\sum_n	\Big[ \;}
		+
 			 \threej{J_b}{0}{J_b}{-m}{0}{m} 
		\frac{
			\bra{J_b} | \gamma_5\rho|  \ket{J_n}  \bra{J_n}  |\hat{d}_{\rm E1}|  \ket{J_a}}
			{E_{b}-E_{n}}
	\Bigg],
\label{eq:si-pnc-rme}
\end{align}
with $m={\rm min}(J_a,J_b)$.

The reduced matrix element of the SI-PNC interaction is defined:
\begin{equation}
\bra{J_a} | \gamma_5\rho|  \ket{J_b} = iR_{SI} C_{SI}, 
\end{equation}
where 
\begin{align}
R_{SI} &= -\alpha\int \rho (f_ag_b -g_af_b)\,{\rm d}r,
\end{align}
is the single-electron radial integral, and $C_{SI} = \sqrt{2J_a+1}$ is the angular coefficient.
(Note that the coefficient $C_{SI}$ and the $3j$-symbol in (\ref{eq:si-pnc-rme}) cancel).

The electron ($F$-independent) part of the SI-PNC amplitude 
(i.e.~with $\ket{a}=\ket{J_a,l_a,m_a}$ in (\ref{eq:fullsummation})) 
is given by the formula

\begin{align}
&E_{\rm PNC} =
\frac{G_F}{2\sqrt{2}}(-Q_W)  
	\sum_n (-1)^{J_b+J_n-2m} \notag\\
	& \times\Bigg[
		\threej{J_b}{1}{J_n}{-m}{0}{m} \threej{J_n}{0}{J_a}{-m}{0}{m}
	\notag\\
		& \phantom{\times\Bigg[} 
		\times \frac{
			\bra{J_b} |\hat{d}_{\rm E1}|  \ket{J_n} \bra{J_n}  |\gamma_5\rho|  \ket{J_a}}
			{E_{a}-E_{n}} \notag\\
		&\phantom{\times\Bigg[} +
		\threej{J_b}{0}{J_n}{-m}{0}{m} \threej{J_n}{1}{J_a}{-m}{0}{m}
\notag\\
		& \phantom{\times\Bigg[} 
		\times
		\frac{
			\bra{J_b} | \gamma_5\rho|  \ket{J_n}  \bra{J_n}  |\hat{d}_{\rm E1}|  \ket{J_a}}
			{E_{b}-E_{n}}
	\Bigg],
\label{eq:fi-pnc-rme}
\end{align}
where for the $z$ component we take $m={\rm min}(J_a,J_b)$.



\begin{thebibliography}{40}
\bibitem{NoteDefinition} 
Note that several different definitions of $\varkappa$ and the SD-PNC Hamiltonian exist in the literature. We have tried to be as consistent as possible, but the reader should be aware that these differences must be taken into account when comparing results.
\end{thebibliography}
\end{document}